\documentclass[10pt,twocolumn,nofootinbib,superscriptaddress]{revtex4}
\usepackage{amssymb}
\usepackage{amsmath}
\usepackage{amsthm}

\usepackage{times}
\usepackage{framed}
\usepackage{enumerate}
\usepackage{hyperref}

\newtheorem{theorem}{Theorem}
\newtheorem{lemma}[]{Lemma}
\newtheorem{remark}[]{Remark}
\newtheorem{definition}[]{Definition}
\theoremstyle{definition}
\newcommand*{\textfrac}[2]{{{#1}/{#2}}}
\newcommand*{\bbR}{\mathbb{R}}

\newcommand*{\cE}{\mathcal{E}}
\newcommand*{\cF}{\mathcal{F}}

\newcommand*{\cH}{\mathcal{H}}

\newcommand*{\cL}{\mathcal{L}}

\newcommand*{\cV}{\mathcal{V}}

\newcommand*{\half}{{\textfrac{1}{2}}}
\newcommand*{\id}{\mathsf{id}}
\newcommand*{\ket}[1]{|#1\rangle}
\newcommand*{\bra}[1]{\langle #1|}
\newcommand*{\proj}[1]{\ket{#1}\bra{#1}}
\newcommand*{\tr}{\mathsf{tr}}
\newcommand*{\rank}{\mathsf{rank}}
\newcommand*{\spr}[2]{\langle #1,#2 \rangle}
\newcommand*{\myspan}{\mathrm{span}}
\newcommand*{\CPTPM}{\mathsf{CPTPM}}
\newcommand*{\CPUM}{\mathsf{CPUM}}

\newcommand*{\hmin}{H_{\min}}
\newcommand*{\hmax}{H_{\max}}
\newcommand*{\hshannon}{S} 
\newcommand*{\hvneu}{S}

\newcommand*{\RelEntr}{D_{\infty}}
\newcommand*{\assign}{\ensuremath{\kern.5ex\raisebox{.05ex}{\mbox{\rm:}}\kern -.3em =}}

\usepackage[caption=false,font=normalsize,labelfont=sf,textfont=sf]{subfig}

\begin{document}

\newcommand*{\myparagraph}[1]{\vspace{0.5em}\paragraph*{{#1}.}}
\renewcommand*{\hat}{\widehat}
\newcommand*{\interior}{\mathsf{int}\ }
\newcommand*{\Herm}{\mathsf{Herm}}

\title{The operational meaning of min- and max-entropy}

\author{Robert K\"onig}
\affiliation{Institute for Quantum Information, California Institute of Technology, Pasadena, USA}
\author{Renato Renner}
\affiliation{Institute for Theoretical Physics, ETH Zurich, Switzerland}
\author{Christian Schaffner}
\affiliation{Centre for Mathematics and Computer Science (CWI), Amsterdam, The Netherlands}

\begin{abstract}
  We show that the conditional min-entropy $\hmin(A|B)$ of a bipartite
  state $\rho_{A B}$ is directly related to the maximum achievable
  overlap with a maximally entangled state if only local actions on
  the $B$-part of $\rho_{A B}$ are allowed. In the special case where
  $A$ is classical, this overlap corresponds to the probability of
  guessing~$A$ given~$B$.  In a similar vein, we connect the conditional
  max-entropy $\hmax(A|B)$ to the maximum fidelity of $\rho_{AB}$ with
  a product state that is completely mixed on~$A$. In the case where
  $A$ is classical, this corresponds to the security of $A$ when used
  as a secret key in the presence of an adversary holding~$B$.  Because min- and max-entropies are known to characterize
  information-processing tasks such as randomness extraction and state
  merging, our results establish a direct connection between these
  tasks and basic operational problems.  For example, they imply that
  the (logarithm of the) probability of guessing~$A$ given $B$ is a
  lower bound on the number of uniform secret bits that can be
  extracted from $A$ relative to an adversary holding~$B$.
\end{abstract}
\maketitle

\section{Introduction}

\newcommand*{\Shannon}{S}
\newcommand*{\eps}{\varepsilon}

\newcommand*{\ellcomp}{\ell_{\mathsf{compr}}}
\newcommand*{\ellext}{\ell_{\mathsf{extr}}}
\newcommand*{\elltransm}{\ell_{\mathsf{transm}}}
\newcommand*{\elldec}{\ell_{\mathsf{decpl}}}
\newcommand*{\ellmerg}{\ell_{\mathsf{merg}}}

\newcommand*{\pguess}{p_{\mathsf{guess}}}
\newcommand*{\psecr}{p_{\mathsf{secr}}}
\newcommand*{\pcorr}{q_\mathsf{corr}}
\newcommand*{\pdecpl}{q_{\mathsf{decpl}}}
\newcommand*{\compressionrate}{r_{\mathsf{compr}}}
\newcommand*{\transmissionrate}{r_{\mathsf{transm}}}

A central goal of information theory is the (quantitative) analysis of
processes involving the acquisition, transmission, and storage of
information. For example, given a (noisy) communication channel, one
may ask at which rate data can be transmitted reliably (this is the
\emph{channel capacity}).  Or, given a source emitting signals, one
may be interested in the amount of space needed to store the
information in such a way that the signal can be recovered later (this
is the \emph{compression rate}).  In the following, we call such
quantities \emph{operational} because they are defined by an actual
information-processing task.

Traditionally, most operational quantities are defined
\emph{asymptotically} under the assumption that a certain process is
repeated many times \emph{independently}.\footnote{The independence
  assumption is sometimes replaced by the less restrictive requirement
  that the process is Markovian.} Consider for
example the problem of data compression. For a random variable $X$ and
for $\eps \geq 0$, let $\ellcomp^\eps(X)$ be the minimum length
(measured in terms of bits) of an encoding $\mathrm{enc}(X)$ such that
$X$ can be recovered from $\mathrm{enc}(X)$ except with an error
probability of at most $\eps$. The \emph{compression rate} of a source
emitting a sequence of mutually independent pieces of data $X_1,
\ldots, X_n$, each distributed according to~$P_X$, is then defined by
\begin{align} \label{eq:compressionrate}
 \compressionrate(P_X) \assign \lim_{\eps \to 0} \lim_{n \to \infty} \frac{\ellcomp^\eps(X_1 \cdots X_n)}{n} \ .
\end{align}

It is maybe one of the most remarkable features of information theory
that a huge variety of operational quantities can be expressed in
terms of a few simple entropy measures. In fact, in the asymptotic
case where a process is repeated many times independently, the
(almost) only relevant entropy measure is the \emph{Shannon entropy}
(or its quantum-mechanical generalization, the \emph{von Neumann
  entropy}). For example, the compression
rate~\eqref{eq:compressionrate} of a source emitting data distributed
according to~$P_X$ is equal to the Shannon entropy $\hshannon$ of a
random variable~$X$ with distribution $P_X$, i.e.,
\begin{align} \label{eq:rateasym}
\compressionrate(P_X) = \hshannon(X) \ .
\end{align} 
This equality is also known as the \emph{source-coding
 theorem}~\cite{Shannon48}.  Another well-known example is the
channel capacity. According to the \emph{noisy-channel coding
 theorem}~\cite{Shannon48}, the maximum rate at which information can
be transmitted over a noisy communication channel is equal to a
difference between two Shannon entropies (see~\eqref{eq:channelcap}
below).

The situation is different in the non-asymptotic case or when the
independence assumption is dropped. Here, the Shannon / von Neumann
entropies no longer give a correct characterization of operational
quantities.\footnote{For example, the minimum compression length
  $\ellcomp^\eps(X)$ defined above can deviate arbitrarily from
  $\hshannon(X)$. This is readily verified by the following example:
  Let $X$ be defined as the random variable which takes the value $0$
  with probability $\frac{1}{2}$, and with probability $\frac{1}{2}$
  is equal to a uniformly distributed bitstring of length $n$.  Then
  $\hshannon(X) \approx \frac{n}{2}$ while $\ellcomp^\eps(X) \approx
  n$ for any sufficiently small~$\eps$.} One therefore has to replace
them by more general entropy measures.  In the past few years, several
such generalizations have been developed, notably the \emph{spectral
  entropy rates}~\cite{VerHan94}, as well as \emph{(smooth) min- and
  max-entropies}~\cite{RenWol04b}.  While both notions completely
overcome the need for independence or Markovian assumptions, spectral
entropy rates are (as suggested by their name) still restricted to
asymptotic considerations. In contrast, smooth min- and max-entropies
are fully general.\footnote{The spectral entropy rates can be seen as
  asymptotic limits of smooth min-/max-entropies~\cite{DatRen08}.}  In
particular, no repetition of random processes is required. That is,
one may consider situations where a source only emits one single piece
of information or where a channel is only used once.

The aim of the present paper is to propose new operational
interpretations of these non-asymptotic entropy measures. Our main
findings are motivated and described in the following subsections,
which are organized as follows.  In Section~\ref{sec:minentropy}, we
review the notion of min-/max-entropies, our central object of
interest. These entropy measures are the basis for the definition of
\emph{smooth} min-/max-entropies, which can be seen as generalizations
of Shannon / von Neumann entropy, as indicated above. Their properties
are discussed later in Section~\ref{sec:minentropy}. After this
preparation, we will turn to connections between (smooth)
min/max-entropies and operational quantities, starting with some
important examples in Section~\ref{sec:opint}. We then summarize the
new operational interpretations derived in this work as well as their
implications in Section~\ref{sec:contribution}.

\subsection{(Smooth) min-/max-entropy: Basic
  definitions\label{sec:minentropy}}

\subsubsection*{Min-/max-entropy\label{sec:minmaxentropy}}

We start with the definition of \emph{conditional min-entropy}. This
quantity (and the closely related conditional max-entropy) is the main
object of study of this paper. In what follows, $\id_A$ denotes the
identity on system $A$.

\begin{definition}
 Let $\rho = \rho_{A B}$ be a bipartite density operator. The
 \emph{min-entropy of $A$ conditioned on $B$} is defined by
 \begin{align} \label{eq:hmindef}
   \hmin(A|B)_{\rho} \assign - \inf_{\sigma_B} \RelEntr(\rho_{A B} \| \id_A \otimes \sigma_B)
 \end{align}
 where the infimum ranges over all normalized density operators $\sigma_B$ on
 subsystem $B$ and where\footnote{For commuting density operators
   $\tau$ and $\tau'$, the quantity $\RelEntr(\tau\|\tau')$
   corresponds to the (classical) relative R\'enyi entropy of order
   $\infty$. In general, the relative R\'enyi entropy of order
   $\alpha$ of two probability distributions $P$ and $Q$ is defined as
   $D_\alpha(P,Q)\assign\frac{1}{\alpha-1}\log_2 \sum_x P_X(x)^\alpha
   Q(x)^{1-\alpha}$, and $D_{\infty}$ is obtained in the limit $\alpha
   \to \infty$. }
 \begin{align} \label{eq:Dinfty} \RelEntr(\tau\|\tau') \assign
   \inf\{\lambda \in \bbR: \, \tau \leq 2^\lambda \tau' \} \ .
 \end{align}
\end{definition}
It is interesting to note that the Shannon / von Neumann entropy could
be defined in a similar way. Namely, if we replace~$\hmin$ by the von Neumann entropy~$\hvneu$ and
$\RelEntr$ by the relative entropy\footnote{Note that the relative  entropy  (aka Kullback-Leibler divergence) $D(\tau \| \tau') \assign \tr(\tau (\log_2 \tau - \log_2  \tau'))$ is also defined 
for unnormalized operators $\tau '$.  }
 $D$ in~\eqref{eq:hmindef}, we find
\begin{align*}
 \hvneu(A|B)_{\rho} = - \inf_{\sigma_B} D(\rho_{A B} \| \id_A \otimes \sigma_B) \ .
\end{align*}
This equality is readily verified using the fact that $D(\tau \|
\tau')$ is nonnegative for any normalized $\tau, \tau'$ and equals
zero if~\mbox{$\tau = \tau'$}.

For a tripartite pure state $\rho = \rho_{A B C}$, the von Neumann
entropy satisfies the equality\footnote{Note that, by definition,
 $\hvneu(A|B) = \hvneu(A B) - \hvneu(B)$ and $\hvneu(A|C) = \hvneu(A C) - \hvneu(C)$. The equality
 then follows from the fact that, by the Schmidt decomposition, $\hvneu(A B) =
 \hvneu(C)$ and $\hvneu(B) = \hvneu(A C)$.}
\begin{align} \label{eq:vNdual}
 \hvneu(A|B)_\rho = -\hvneu(A|C)_{\rho} \ .
\end{align}
The same is no longer true for the min-entropy. However, it
turns out that the entropy obtained by replacing the system $B$ by the
``purifying system''~$C$ often appears in expressions characterizing
operational quantities. This motivates the following definition.
\begin{definition}\footnote{In the existing literature, $H_{\max}$ and $H_{\max}^{\varepsilon}$ are sometimes defined in a different manner (closely related to the R\'enyi entropy of order~$0$). It can be shown, however, that the smooth variants of these definitions only deviate by an additive term which is logarithmic in the smoothness parameter (see~\cite{ToCoRe08}).}
  Let $\rho = \rho_{A B}$ be a bipartite density operator. The
  \emph{max-entropy of $A$ conditioned on $B$} is defined by
 \begin{align}\label{eq:purificationdef}
   \hmax(A | B)_{\rho} \assign - \hmin(A | C)_{\rho}
 \end{align}
 where the min-entropy on the rhs.\ is evaluated for a
 purification $\rho_{A B C}$ of $\rho_{A B}$.
\end{definition}
\noindent This is well-defined because all purifications of
$\rho_{AB}$ are related by unitaries on~$C$, and the
quantity~$\hmin(A|C)_\rho$ is invariant under such unitaries.

We point out that $\hmin$ and $\hmax$ could have been defined
alternatively by starting from an expression for $\hmax$ and
subsequent definition of $\hmin$ by purification
(i.e.,~\eqref{eq:purificationdef}). In this sense, both quantities are
equally fundamental.

If the state $\rho$ is clear from the context, we will omit the
subscript in $\hmin(A|B)_{\rho}$ and $\hmax(A | B)_{\rho}$.
Also, in the special case where the system $B$ is trivial (i.e.,
one-dimensional), we omit the conditioning and simply write
$\hmin(A)$ and $\hmax(A)$. Note that the above definitions also
apply to classical probability distributions $P_X$ which can always be
written as quantum states $\rho_{X} = \sum_{x} P_X(x) \proj{x}$ for
some orthonormal basis $\{\ket{x}\}_{x}$.

To get some more intuition for these definitions, it may help to
compute their value for certain special states. One extreme case are
product states $\rho_{A B} = \rho_A \otimes \rho_B$, for which one
readily verifies that the min-entropy only depends on the maximum
eigenvalue $\| \rho_A \|_{\infty}$ of $\rho_A$, i.e., $\hmin(A|B)_\rho
= -\log_2 \| \rho_A \|_{\infty}$. Note that this corresponds to the
R\'enyi entropy of order infinity of the density operator $\rho_A$.
Similarly, we get $\hmax(A|B)_\rho=2 \log_2 \tr\sqrt{\rho_A}$, which
is the R\'enyi entropy of order~$\frac{1}{2}$ of $\rho_A$ (see
eq.~\eqref{eq:hmaxproduct} below).  Another extreme case is where
$\rho_{A B}$ is a pure state. Here one finds $\hmin(A|B)_\rho = -
\log_2 (\tr\sqrt{\rho_A})^2$ and \mbox{$\hmax(A|B)_\rho=\log_2
\|\rho_A\|_\infty$}.

\subsubsection*{Smooth min-/max-entropy}

The \emph{smooth} min/max-entropy of a state $\rho$ is defined by the
corresponding (non-smooth) min/max-entropy for an ``optimal'' state
$\rho'$ in an $\eps$-neighborhood of $\rho$, where $\eps$ is called
\emph{smoothness parameter}.

\begin{definition}
 Let $\rho = \rho_{A B}$ be a bipartite density operator and let
 $\eps \geq 0$. The \emph{$\eps$-smooth min- and max-entropy of $A$
   conditioned on $B$} are given by
 \begin{align*}
   \hmin^\eps(A | B)_\rho &\assign \sup_{\rho'}
   \hmin(A|B)_{\rho'} \, ,\\
   \hmax^\eps(A | B)_\rho &\assign \inf_{\rho'} \hmax(A|B)_{\rho'}\ .
 \end{align*}
 where the supremum ranges over all density operators~\mbox{$\rho'=\rho'_{AB}$} which are $\eps$-close to $\rho$.\footnote{In the classical
   case, smooth entropies are usually defined with respect to the
   trace distance
   $\delta_{\tr}(\rho,\sigma)=\frac{1}{2}\|\rho-\sigma\|_1$. Quantum-mechanically,
   distance measures based on the fidelity~$F(\rho, \sigma) = \|
   \sqrt{\rho} \sqrt{\sigma} \|_1$ are more suitable because they
   are invariant under purifications. Candidates are the {\em Bures
     distance } $\|\rho-\sigma\|_B=\sqrt{2-2F(\rho,\sigma)}$
   and the {\em angle} $\|\rho-\sigma\|_A =\arccos F(\rho,\sigma)$. The corresponding definitions are
   essentially equivalent because of the inequalities
   $1-F(\rho,\sigma)\leq \delta_{\tr}(\rho,\sigma) \leq
   \sqrt{1-F(\rho,\sigma)^2}$.}
\end{definition}

\myparagraph{Basic properties}

It follows directly from the definitions that the same kind of duality
between min- and max-entropy holds between the corresponding smooth
versions, namely
\begin{align} \label{eq:hminmaxeps}
  \hmax^\eps(A | B)_{\rho} = - \hmin^\eps(A | C)_{\rho}
\end{align}
for a purification $\rho_{A B C}$ of $\rho_{A B}$.

As already indicated, smooth min-/max-entropies can be seen as
generalizations of the Shannon / von Neumann entropy $\hshannon$. More
precisely, the latter can be written in terms of the
former~\cite{Ren05,ToCoRe08}, i.e.,
\begin{align} \label{eq:asym}
 \hvneu(A|B)_{\rho} 
& = 
 \lim_{\eps \to 0} \lim_{n \to \infty} \frac{1}{n} \hmin^\eps(A^n | B^n)_{\rho^{\otimes n}} \\ \label{eq:asymmax}
 \hvneu(A|B)_{\rho} 
& = 
 \lim_{\eps \to 0} \lim_{n \to \infty} \frac{1}{n} \hmax^\eps(A^n | B^n)_{\rho^{\otimes n}}
\end{align}
Note that the two statements are trivially equivalent because
of~\eqref{eq:vNdual} and~\eqref{eq:hminmaxeps}.

Given these asymptotic relations, it is not surprising that smooth
min-/max-entropies share various properties with the Shannon / von
Neumann entropy. For example, they are strongly subadditive, i.e.,
\begin{align} \label{eq:strongsub}
 \hmin^\eps(A|B) \geq \hmin^\eps(A | B C)
\end{align}
and likewise for $\hmax^\eps$. In fact,
inequality~\eqref{eq:strongsub} can be seen as a generalization of the
strong subadditivity of the von Neumann entropy, $\hvneu(A|B) \geq
\hvneu(A | B C)$, which can be recovered by virtue of
identity~\eqref{eq:asym}, i.e., for any $\rho_{A B C}$,
\begin{align*}
 \hvneu(A | B)_{\rho} 
&\stackrel{\eqref{eq:asym}}{=}
 \lim_{\eps \to 0} \lim_{n \to \infty} \frac{1}{n} \hmin^\eps(A^n | B^n)_{\rho^{\otimes n}}\\
&\stackrel{\eqref{eq:strongsub}}{\geq}
 \lim_{\eps \to 0} \lim_{n \to \infty} \frac{1}{n} \hmin^\eps(A^n | B^n C^n)_{\rho^{\otimes n}}\\
&\stackrel{\eqref{eq:asym}}{=}
 \hvneu(A | B C)_{\rho} \ .
\end{align*}
Interestingly, despite its generality, inequality~\eqref{eq:strongsub}
is easy to prove, as we shall see at the end of
Section~\ref{sec:contribution}.

\subsection{Operational quantities in terms of smooth
  min-/max-entropy} \label{sec:opint}

The main reason for considering (smooth) min-/max-entropies is that
they are well suited for the characterization of operational
quantities in the most general case.  Recall that expressions for
operational quantities involving the Shannon / von Neumann entropy,
e.g., \eqref{eq:rateasym}, are typically only valid asymptotically,
under the assumption that certain resources can be used many times
independently. Interestingly, the structure of such expressions
essentially remains the same if one drops these assumptions, except
that smooth entropies take the place of Shannon / von Neumann
entropy. The purpose of this section is to illustrate this phenomenon
with a few examples.

\myparagraph{Data compression}

We start with the example of data compression, which has already been
introduced above. For a random variable $X$ and $\eps \geq 0$, let
again $\ellcomp^\eps(X)$ be the minimum length of an encoding from
which the value of $X$ can be recovered correctly with probability at
least $1-\eps$.  It can then be shown that $\ellcomp^\eps(X)$ is
essentially equal to the smooth max-entropy of $X$~\cite{RenWol04b}. More
precisely, we have
\begin{align} \label{eq:compression}
\ellcomp^{\eps}(X) 
=
 \hmax^{\eps'}(X) + O(\log\textfrac{1}{\eps})
\end{align}
for some $\eps' \in [\frac{1}{2} \eps, 2 \eps]$. The $O$-notation
indicates that equality holds up to an additive term of the order
$\log(\textfrac{1}{\eps})$.\footnote{Note that the smooth entropies
  are monotonic functions of $\eps$. Equality~\eqref{eq:compression}
  is thus just a way to state that the operational quantity
  $\ellcomp^{\eps}(X)$ lies in the interval $[ \hmax^{2\eps}(X),\hmax^{\eps/2}(X)]$, up to some additive constant of the order~$\log\textfrac{1}{\eps}$.}  In typical applications, this
logarithmic term is much smaller than the other quantities occurring
in the expression. In particular, the term is independent of the size
of the resource (in our case the random variable $X$) and thus becomes
irrelevant in the asymptotic limit of large resources.

We stress that~\eqref{eq:compression} is valid for a \emph{single}
realization of the random variable $X$ and thus strictly generalizes
Shannon's source coding theorem described at the beginning of this
section. Identity~\eqref{eq:rateasym} can be recovered as an
asymptotic limit of~\eqref{eq:compression}, for $X$ consisting of many
independent and identically distributed pieces $X_1, \ldots, X_n$,
i.e.,
\begin{align*}
 \compressionrate(P_X) 
&\stackrel{\eqref{eq:compressionrate}}{=}
 \lim_{\eps \to 0} \lim_{n \to \infty} \frac{\ellcomp^\eps(X_1 \cdots X_n)}{n}\\
&\stackrel{\eqref{eq:compression}}{=}
 \lim_{\eps \to 0} \lim_{n \to \infty} \frac{1}{n} \hmax^\eps(X_1 \cdots X_n)\\
&\stackrel{\eqref{eq:asymmax}}{=}
 \hshannon(X)  \ .
\end{align*}

\myparagraph{Channel coding}

As a second example, we consider the noisy-channel coding
problem. For any $\eps \geq 0$, let $\elltransm^\eps(X \to Y)$ be the
maximum number of bits that can be transmitted in \emph{one use} of a
classical noisy channel $X \to Y$ (specified by a conditional
probability distribution $P_{Y|X}$) with maximum error probability
$\eps$. As shown in~\cite{ReWoWu06}, this quantity is given by
\begin{align} \label{eq:onecap}
\begin{split} \elltransm^{\eps}(X\to Y) = \max_{P_X} \bigl( \hmin^{\eps'}(X) - \hmax^{\eps'}(X|Y) \bigr) \\
+ O(\log \textfrac{1}{\eps})
\end{split}
\end{align}
for some $\eps' \in [\frac{1}{2} \eps, 2 \eps]$.

Similarly to the example of source coding, we may consider the special
case where the channel allows many mutually independent
transmissions. The figure of merit then is the \emph{channel
  capacity}
\begin{align*}
  \transmissionrate(P_{Y|X}) 
\assign 
  \lim_{\eps \to 0} \lim_{n \to \infty} \frac{\elltransm^\eps(X^n \to Y^n)}{n} \ , 
\end{align*}
that is, the maximum rate at which information can be transmitted by
$n$ uses of the channel $X \to Y$, in the limit of large $n$. Using
the non-asymptotic statement~\eqref{eq:onecap} together
with~\eqref{eq:asym} and~\eqref{eq:asymmax}, we find
\begin{align} \label{eq:channelcap}
 \transmissionrate(P_{Y|X}) 
=
 \max_{P_X} \bigl( \hshannon(X) - \hshannon(X|Y) \bigr)
= 
 \max_{P_X} I(X:Y) \ .
\end{align}
This is Shannon's well-known noisy-channel coding theorem.

\myparagraph{Privacy amplification}

Let $X$ be a classical random variable and let $B$ be (possibly
quantum-mechanical) side information. The goal of \emph{randomness
  extraction} is to compute a bitstring $f(X)$ which is uniform and
independent of the side information $B$.  Randomness extraction is
crucial for a number of applications, particularly in the context of
cryptography, where it is also called \emph{privacy
  amplification}~\cite{BBCM95}. For example, in a key-agreement
scheme, one may want to turn a (only partially secure) raw key $X$
into a fully secure key~$f(X)$. Security of $f(X)$ is then akin to
uniformity relative to side information $B$ held by a potential
adversary.

The maximum number of uniform and independent bits that can be
extracted from $X$ is directly given by the smooth min-entropy of
$X$. More precisely, let $\ellext^\eps(X|B)$ be the maximum length of
a bitstring $f(X)$ that can be computed from $X$ such that $f(X)$ is
$\eps$-close to a string $Z$ which is perfectly uniform and
independent of the side information~$B$.\footnote{See paragraph on
  max-entropy of classical information in
  Section~\ref{sec:contribution} below for more details.} One can show
that~\cite{Ren05,KoeRen08}
\begin{align*}
 \ellext^\eps(X|B) = \hmin^{\eps'}(X|B) + O(\log\textfrac{1}{\eps})
\end{align*}
where $\eps' \in [\frac{1}{2} \eps, 2 \eps]$. In the special case
where $B$ is independent of $X$, this corresponds to the
\emph{leftover hash lemma}~\cite{ILL89,BBCM95}. For later
reference, we also note that
\begin{align} \label{eq:PAlowerbound}
 \ellext^\eps(X|B) \geq \hmin(X|B) + O(\log\textfrac{1}{\eps}) \ ,
\end{align}
which holds because $\hmin^\eps(X|B)$ is monotonically increasing in
$\eps$ and equals $\hmin(X|B)$ for $\eps = 0$.

\myparagraph{Decoupling}

The previous result can be extended to a fully quantum-mechanical
setting as follows. Let $A$ and $B$ be two quantum systems. The goal
is to find a maximum subsystem~$A'$ of~$A$ such that the state on $A'$
is completely mixed and decoupled from $B$ (conditioned on a suitable
measurement on the remaining part of $A$).  Let $\elldec^{\eps}(A|B)$
be the maximum size of $A'$ (measured in qubits) such that this is
possible up to a distance~$\eps$.\footnote{This distance is quantitatively
  expressed by the {\em decoupling accuracy}, see below.} One then
finds~\cite{horodeckietal05,WinRen07,Berta08}\footnote{This is based on a tightened version~\cite{WinRen07} of
  a bound obtained in~\cite{horodeckietal05}, which shows that
  projecting onto a random subspace of dimension $\dim A'$ achieves
  decoupling. More precisely, it can be shown~\cite{WinRen07} that the
  decoupling accuracy of the residual state is, on average over the
  measurement outcome, exponentially small in the difference
  $\hmin(A|B)-\log \dim A'$.}
\begin{align} \label{eq:elldecpl}
 \elldec^\eps(A|B) = \hmin^{\eps'}(A|B) + O(\log\textfrac{1}{\eps}) \ 
\end{align}
with $\varepsilon'\in [\frac{1}{2}\varepsilon,2\varepsilon]$.

\myparagraph{State merging}

In the same manner as privacy amplification generalizes to decoupling
in the fully quantum case, data compression (and its relatives such as
coding with side information) extends to a fully quantum setting; this
is referred to as state merging. The setting is described by a
tripartite pure state $\ket{\Psi_{ABC}}$. The aim is to redistribute
the $A$-part to the system $B$ by local operations and classical
communications (LOCC) between $A$ and $B$. Depending on the (reduced)
state $\rho = \rho_{AB}$, this either consumes or generates bipartite
entanglement.  Let (-)$\ellmerg^\eps(A|B)_{\rho}$ be the minimal
(maximal) number of ebits of entanglement required (generated) by this
process (the distinction between consumed/generated entanglement is
reflected by the sign of the quantity~$\ellmerg^\eps(A|B)_{\rho}$),
such that the outcome is $\varepsilon$-close to the desired
output.\footnote{Closeness is measured in terms of the distance of the
  output state $\rho_{ABB_1B_2C}'$ of the protocol to the state
  $\ket{\Phi_{AB}}^{\otimes \ell}\otimes \ket{\Psi_{B_1B_2C}}$, where
  $\ket{\Phi_{AB}}$ is an ebit between $A$ and $B$, $\ell$ is the
  number of ebits generated, and $\ket{\Psi_{B_1 B_2 C}}$ is identical
  to $\ket{\Psi_{ABC}}$ when identifying the subsystems $B_1$ and $A$
  as well as $B_2$ and $B$.}  One then finds
\begin{align} \label{eq:ellmerg}
 \ellmerg^\eps(A|B)_\rho = \hmax^{\eps'}(A|B)_\rho + O(\log\textfrac{1}{\eps}) \ ,
\end{align}
where again~$\varepsilon'\in [\frac{1}{2}\varepsilon,2\varepsilon]$
(see~\cite{Berta08} for details). In fact, the $\geq$-part of this
statement is a direct consequence of decoupling result
above~\cite{WinRen07}, the arguments in~\cite{horodeckietal05}
(cf.~also Section~\ref{sec:proofhmax}) and the definition of
$\hmax(A|B)_\rho$.

\subsection{Contribution: Min-/max-entropies as operational
  quantities\label{sec:contribution}}

In this paper, we show that min-/max-entropies have
direct\footnote{The term \emph{direct} refers to the fact that no
  smoothing is required, in contrast to the examples of
  Section~\ref{sec:opint}.} operational interpretations. We begin by
presenting the corresponding results for the special case where we
condition classical information~$X$ on a (possibly) quantum system
$B$. The fully general case is discussed later in
Section~\ref{sec:quantumresults}.

\subsubsection{Uncertainty about classical information}

Consider an agent with access to a (classical or quantum) system $B$
whose state $\rho_B^x$ depends on a classical random variable
$X$. This situation can be described by a classical-quantum state
\begin{align} \label{eq:cqstate}
 \rho = \rho_{X B} 
\assign
 \sum_x P_X(x) \proj{x} \otimes \rho_B^x \ ,
\end{align}
with $\{\ket{x}\}_x$ a family of mutually orthogonal vectors
representing the (classical) values of $X$.

\myparagraph{Min-entropy of classical information is guessing
  probability}

Let $\pguess(X|B)$ be the probability that the agent correctly guesses
$X$ when using an optimal strategy; that is, $\pguess(X|B)=\sum_x
P_X(x)\tr(E_x\rho_B^x)$, where the optimal measurement strategy is
described by the POVM~$\{E_x\}_x$ on $B$ that maximizes this
expression. Note that conditions for the optimality of a POVM~$\{E_x\}_x$ in this hypothesis testing problem were found by Holevo~\cite{Holevo73} and independently by Yuen, Kennedy and Lax~\cite{Yuenetal75}. These works also use semidefinite programming duality in a similar fashion as in this paper. Here we are interested in the optimal value of this optimization problem.  We show that (cf.~Theorem~\ref{lem:hguess})
\begin{align} \label{eq:guessop}
 \pguess(X|B) = 2^{-\hmin(X|B)_{\rho}} \ .
\end{align}
where the entropy is evaluated for the state~$\rho_{X B}$ given
by~\eqref{eq:cqstate}.

If no side information $B$ is available or, more generally, if the
state of $B$ is independent of $X$, we have $2^{-\hmin(X|B)} = \|
\rho_X\|_{\infty} = \max_x P_X(x)$ as noted in
Section~\ref{sec:minentropy}. Identity~\eqref{eq:guessop} then reduces
to the trivial fact that the maximum probability of
correctly guessing~$X$ without prior information is equal to $\max_x
P_X(x)$.

Note that previously, only the upper bound~\cite{christandl07}
\begin{align*}
 \pguess(X|B)\leq 2^{-\hmin(X|B)_{\rho}} \ 
\end{align*}
and the lower bound~\cite{christandletal06}
\begin{align*}
2^{-H_2(X|B)_{\rho}}\leq \pguess(X|B)\ 
\end{align*}
were known, where the lhs.~is the average guessing probability when the square-root measurement~\cite{hausladenetal96} is used, that is,
$H_2(X|B)_{\rho}=-\log \tr\left(\left((\id_X\otimes\rho^{-\textfrac{1}{2}}_B)\rho_{XB}\right)^2\right)$.

\myparagraph{Max-entropy of classical information is security of key}

The secrecy of $X$ when used as a key in the presence of an adversary
with access to system $B$ is conventionally measured in terms of the
distance of the state $\rho_{X B}$ (cf.~\eqref{eq:cqstate}) to a
product state of the form $\tau_X \otimes \rho_B$, where $\tau_X$ is
the completely mixed state (corresponding to the uniform distribution
on $X$) and where $\rho_B$ is the reduced state on subsystem $B$. This
models an \emph{ideal} situation where the key is perfectly uniform
and independent of the adversary's system. If the trace distance is
used, then this distance is directly related to the {\em
  distinguishing advantage} between the real and the ideal system.

One may relax the above and only require that the desired state
is of the form $\tau_X \otimes \sigma_B$, for some \emph{arbitrary}
density operator $\sigma_B$. When using the trace distance, this
relaxed definition is equivalent to the above up to a factor of
$2$. Also, since the trace distance and the fidelity are essentially
equivalent, we can use the fidelity. We
then get the following measure for the secrecy of $X$ relative to $B$,
\begin{align*}
\psecr(X|B)_\rho& \assign |X|\max_{\sigma_B} F(\rho_{X B},\tau_X\otimes\sigma_B)^2\\
&=\max_{\sigma_B} \left(\sum_x \sqrt{P_X(x)} F(\rho^x_B,\sigma_B)\right)^2\ .
\end{align*}
where $|X|$ is the alphabet size of $X$ (we include this factor here
for convenience). We show that
(cf.~Theorem~\ref{thm:hmaxinterpretation})
\begin{align}
\psecr(X|B)_\rho &=2^{\hmax(X|B)_\rho}\ .\label{eq:keysecurity}
\end{align}
If no side information $B$ is available or, more generally, if $B$ is
independent of $X$, we obtain \mbox{$2^{\hmax(X|B)_\rho}=\left(\sum_x
  \sqrt{P_X(x)}\right)^2$}
(cf.~Section~\ref{sec:minentropy}). Identity~\eqref{eq:keysecurity}
then simply expresses the fact that the secrecy of $X$ in this case is
quantified by the distance of $P_X$ to the uniform distribution (where
distance is measured in terms of the fidelity).

\subsubsection{Uncertainty about quantum information}
\label{sec:quantumresults}

We now discuss the fully general case, where we have an arbitrary
bipartite state $\rho=\rho_{AB}$. The min-/max-entropies carry the
following operational interpretations.

\myparagraph{Min-entropy is maximum achievable singlet fraction}

Define the maximally entangled state
\begin{align*}
 \ket{\Phi_{A B}} \assign \frac{1}{\sqrt{d}} \sum_{x} \ket{x_A} \ket{x_B}
\end{align*}
where $\{\ket{x_A}\}_{x=1}^{d}$ is an orthonormal basis of subsystem
$A$ (of dimension $d$) and $\{\ket{x_B}\}_{x=1}^d$ is a family of
mutually orthogonal vectors on subsystem $B$ (we assume that $\dim
A\leq \dim B$). We define the ``quantum correlation''
$\pcorr(A|B)_{\rho}$ as the maximum overlap with the singlet\footnote{In the literature, the expression ``singlet''  often refers to the maximally entangled two-qubit state~$\frac{1}{\sqrt{2}}(\ket{0}\ket{1}-\ket{1}\ket{0})$. Here we use the expressions ``singlet'' and ``singlet fraction'' more generally for any maximally entangled state~$\ket{\Phi_{AB}}$.  This is justified because definition~\eqref{eq:pcorrdef} gives the same value independent of the choice of the maximally entangled state~$\ket{\Phi_{AB}}$.} state
$\ket{\Phi_{A B}}$ that can be achieved by local quantum operations
$\cE$ (trace-preserving completely positive maps) on subsystem~$B$,
that is
\begin{align} \label{eq:pcorrdef}
 \pcorr(A|B)_{\rho} \assign d \max_{\cE} F\bigl((\id_A \otimes \cE)(\rho_{A
   B}), \proj{\Phi_{AB}}\bigr)^2\ .
\end{align}
We show that (cf.~Theorem~\ref{thm:singletfraction})
\begin{align}\label{eq:pcorrelation}
 \pcorr(A|B)_\rho = 2^{-\hmin(A|B)_\rho}\ .
\end{align}

Note that in the case where the information is classical, i.e., if
$\rho=\rho_{X B}$ is of the form~\eqref{eq:cqstate}, we have
\begin{align*}
 \pcorr(X|B)_{\rho}
=
 \max_{\cE} \sum_x P_X(x) \bra{x} \cE(\rho_B^x) \ket{x} \ .
\end{align*}
The operation $\cE$ can be interpreted as a guessing strategy, so that
$\bra{x} \cE(\rho_B^x) \ket{x}$ becomes the probability of correctly
guessing $X$ if $X=x$. We thus recover the maximum guessing
probability $\pguess$ as a special case, i.e.,
\begin{align*}
 \pcorr(X|B)_{\rho} = \pguess(X|B) \ . 
\end{align*}

\myparagraph{Max-entropy is decoupling accuracy} 

The \emph{decoupling accuracy} is a parameter that can be seen as the
quantum analogue of the error probability in classical coding theorems
and is also called {\em quantum error}
in~\cite{horodeckietal05,horodeckietal08}; it measures the quality of
decoupling as follows. It is defined as the distance of~$\rho_{AB}$ to
the product state $\tau_A\otimes\sigma_B$, where $\tau_A$ is the
completely mixed state on $A$ and $\sigma_B$ is an arbitrary density
operator. In a cryptographic setting, it quantifies how random $A$
appears from the point of view of an adversary with access to $B$.  As
above for classical $A$, we define a fidelity-based version of this
quantity as
\begin{align} \label{eq:pdecpldef}
 \pdecpl(A|B)_{\rho} \assign d_A \max_{\sigma_B} F(\rho_{AB},\tau_A\otimes\sigma_B)^2\ ,
\end{align}
where  $d_A$ is the dimension of~$A$ and $\tau_A$ is the
completely mixed state on $A$. 
We show that (cf.~Theorem~\ref{thm:hmaxinterpretation})
\begin{align}\label{eq:qerr}
\pdecpl(A|B)_{\rho}&=2^{\hmax(A|B)_\rho}\ .
\end{align}
It is immediately obvious that this generalizes the security parameter
for a classical key $X$, i.e., for $\rho=\rho_{X B}$ of the
form~\eqref{eq:cqstate}, we have
\begin{align*}
\pdecpl(X|B)_{\rho}&=\psecr(X|B)_\rho\ .
\end{align*}

\myparagraph{Implications}

A main implication of our results is that they establish a connection
between seemingly different operational quantities. For example,
because the number $\ellext^\eps(X|B)$ of uniform bits that can be
extracted from $X$ with respect to side information $B$ is lower
bounded by $\hmin(X|B)$ (see~\eqref{eq:PAlowerbound}), we find that
\begin{align*}
 \ellext^\eps(X|B) \geq - \log_2 \pguess(X|B) + O(\log\textfrac{1}{\eps}) \ ,
\end{align*}
In other words, the negative logarithm of the guessing probability of
$X$ tells us how many uniform bits we can extract from $X$ (relative
to some system $B$). This connection between randomness extraction and
guessing entropy may be useful for applications, e.g., in
cryptography. Here, the derivation of lower bounds on the amount of
extractable randomness is usually a central part of the security
analysis (see~\cite{Ren05,DFRSS07,WST08,STW08}).

Our results can also be used to prove additivity properties of the
min-/max-entropies. One of them is additivity of the
min-/max-entropies for independent systems.  Let \mbox{$\rho_{A A' B B'} =
\rho_{A B} \otimes \rho_{A' B'}$}. Then, by the definition of $\pcorr$,
\begin{align*}
  \pcorr(A A' | B B') \geq \pcorr(A|B) \cdot \pcorr(A'|B') \ .
\end{align*}
By virtue of~\eqref{eq:pcorrelation}, this is equivalent to
\begin{align*}
  \hmin(A A' | B B') \leq \hmin(A|B) + \hmin(A'|B') \ .
\end{align*}
Note that the opposite inequality follows immediately from the
definition of $\hmin$ and the additivity of $\RelEntr$. We thus
have
\begin{align*}
  \hmin(A A' | B B')_{\rho} & = \hmin(A|B) + \hmin(A'|B') 
\intertext{and, equivalently (by the definition~\eqref{eq:purificationdef})}
  \hmax(A A' | B B')_{\rho} & = \hmax(A|B) + \hmax(A'|B')
\end{align*}

A second example is the strong subadditivity of conditional
min-entropy~\eqref{eq:strongsub}. Here, it suffices to notice that
every trace-preserving completely positive map $\cE$ acting on $B$ can
also be understood as acting on registers $B$ and $C$, hence, 
\begin{align*} 
  \pcorr(A|B)_\rho \leq \pcorr(A|BC)_\rho
\end{align*}
for every quantum state $\rho_{ABC}$. By~\eqref{eq:pcorrelation}, this
is equivalent to
\begin{align*}
  \hmin(A|B)_\rho \geq \hmin(A|BC)_\rho
\end{align*}
The extension to smooth min-entropy~\eqref{eq:strongsub} is
straightforward (see Lemma 3.2.7 of~\cite{Ren05}).

Our results also simplify the calculation of the
min-/max-entropies. As an example, let us calculate the entropy
$\hmax(A|B)_{\rho}$ for a state of the form $\rho_{A B} = \rho_A
\otimes \rho_B$. By~\eqref{eq:qerr}, it suffices to determine the
quantity $\pdecpl(A|B)_{\rho}$, which is given by
\begin{align*}
  \pdecpl(A|B)_{\rho}
=
  \max_{\sigma_B} d_A F(\rho_A \otimes \rho_B, \tau_A \otimes \sigma_B)
\end{align*}
where $\tau_A$ is the completely mixed state on the $d_A$-dimensional
Hilbert space $A$. Using the multiplicativity of the fidelity, we
find
\begin{align*}
  \pdecpl(A|B)_{\rho}
&=
  d_A F(\rho_A,\tau_A) \max_{\sigma_B} F(\rho_B,\sigma_B)\\
&=
  d_A F(\rho_A,\tau_A)\\
&=
  \| \sqrt{\rho_A} \|_1^2 \ .
\end{align*}
We thus obtain
\begin{align} \label{eq:hmaxproduct}
  \hmax(A|B)_{\rho} 
=
  2 \log \tr \sqrt{\rho_A} \ ,
\end{align}
for any $\rho = \rho_{A B}$ of the form $\rho_{A} \otimes \rho_{B}$. This corresponds to the R\'enyi entropy of order $\frac12$, which is
hence the natural counterpart to the min-entropy (R\'enyi entropy of
order $\infty$). As noted in~\cite{RenWol04b}, the R\'enyi entropy of
order $\alpha$, for any $\alpha < 1$, is --up to small additive terms of
the order $\log\frac{1}{\eps}$-- determined by a smoothed version of
$H_0(\rho_A) \assign \log_2 \rank(\rho_A)$. The max-entropy $H_{\max}$
of a density operator can thus be interpreted as a measure for its
rank.

\subsubsection*{Outline of the remainder of this paper}

In Section~\ref{sec:preliminaries}, we discuss some mathematical
preliminaries, in particular semidefinite programming, which plays a
crucial role in our arguments. Our main results are then stated and
proved in Section~\ref{sec:mainresultproof}.

\section{Some technical preliminaries~\label{sec:preliminaries}}

\subsection{Semidefinite programming\label{sec:semidefinite}}

Our central tool will be the duality between certain pairs of
semidefinite programs. It will be convenient to use a fairly general
formulation of this duality; a derivation of the results summarized in
this section can be found, e.g., in~\cite[Section~6]{barvinok02}. The
presentation here follows this reference, but specializes certain
statements to the situation of interest for simplicity. We start by
introducing a few definitions.

A subset $K\subset\cV$ of a vector space~$\cV$ is called a {\em convex
  cone} if $0\in K$ and $\mu v+\nu w \in K$ for all nonnegative
$\mu,\nu\geq 0$ and $v,w \in K$. A convex cone $K$ gives rise to a
partial order relation~$\leq_K$ on~$\cV$, defined by $v\leq_K w$ if
and only if $w-v\in K$. If $\cV$ is a Euclidean space with inner
product~$\spr{\cdot}{\cdot}$, then the {\em dual cone} $K^*\subset\cV$
of $K$ is defined by \mbox{$K^*=\{v\in\cV\ |\ \spr{v}{w}\geq 0\textrm{ for
  all }w\in K\}$}. The {\em interior} $\interior K\subset K$ is the
subset of points $w\in K$ for which there exists an open ball centered
around~$w$ and contained in~$K$.

Let $\cV_1$ and $\cV_2$ be Euclidean spaces with inner
products~$\spr{\cdot}{\cdot}_1$ and $\spr{\cdot}{\cdot}_2$,
respectively. A linear map \mbox{$E^{*}:\cV_2\rightarrow\cV_1$} is called
{\em dual of} or {\em adjoint to} a linear map
$E:\cV_1\rightarrow\cV_2$ if
\begin{align*}
\spr{E v_1}{v_2}_2&=\spr{v_1}{E^*v_2}_1\qquad\textrm{ for all }v_1\in\cV_1, v_2\in\cV_2\ .
\end{align*}
For a given map $E$, the dual map $E^*$ is necessarily unique if it
exists.  The two linear programming problems we are interested in are
defined in terms of a pair of such maps. They are referred to as the
{\em primal} and {\em dual} problem, and are specified by parameters
$c\in \cV_1$ and $b\in \cV_2$. The programs are expressed by the following
optimizations:
\begin{align}
\begin{matrix}
\gamma^{\mathsf{primal}}&=&\inf_{\substack{
v_1\geq_{K_1} 0\\
E v_1\geq_{K_2} b
}} \spr{v_1}{c}_1 \  ,\\
\gamma^{\mathsf{dual}}&=&\sup_{\substack{
v_2\geq_{K_2^*} 0\\
E^*v_2\leq_{K_1^*} c
}} \spr{b}{v_2}_2\  .
\end{matrix}\label{eq:primalanddual}
\end{align}
We will usually assume that the sets we optimize over are non-empty. (In the language of linear programming, there exists a {\em feasible plan} and a {\em dual feasible plan}.) The {\em weak duality theorem} states that $\gamma^{\mathsf{primal}}\geq \gamma^{\mathsf{dual}}$. We are particularly interested in conditions for equality. (This is referred to as a {\em zero duality gap}.) A simple criterion is {\em Slater's interiority condition}, which states the following 
\begin{lemma}\label{lem:slater}
  Suppose that there is an element $v\in \interior K_1$ such that $E
  v-b\in\interior K_2$. Suppose further that the infimum
  in~\eqref{eq:primalanddual} is attained. Then
  $\gamma^{\mathsf{primal}}=\gamma^{\mathsf{dual}}$.
\end{lemma}

\subsection{Quantum operations}

Let $\cH_A$ be a Hilbert space and let $\cL(\cH_A)$ be the set of
linear maps $E :\cH_A\rightarrow\cH_A$. An element $E \in \cL(\cH_A)$
is called {\em nonnegative} (written $E\geq 0$) if
$\bra{\psi}E\ket{\psi}\geq 0$ for all $\ket{\psi}\in\cH_A$. A {\em
  positive} element $E$ (written $E>0$) is defined in the same way
with a strict inequality.

An {\em operation} is a linear map $\cE:\cL(\cH_A)\rightarrow
\cL(\cH_B)$. It is called {\em trace-preserving} if
$\tr(\cE(E))=\tr(E)$ for all $E\in\cL(\cH_A)$. It is {\em unital} if
it maps the identity on $\cH_A$ to the identity on $\cH_B$, i.e., if
$\cE(\id_{A})=\id_B$. The map is called {\em positive} if $\cE(E)
\geq 0$ for all $E \geq 0$. It is {\em completely positive} (CP) if
$\id_R \otimes
\cE:\cL(\cH_R\otimes\cH_A)\rightarrow\cL(\cH_R\otimes\cH_B)$ is
positive for any auxiliary space~$\cH_R$, where $\id_R$
is the \emph{identity operation}. A {\em quantum operation} is a
completely positive trace-preserving map (CPTP).  The {\em adjoint
  map} of an operation $\cE:\cL(\cH_A)\rightarrow\cL(\cH_B)$ is the
unique map $\cE^\dagger:\cL(\cH_B)\rightarrow\cL(\cH_A)$ satisfying
\begin{align*}
\tr(F_B\cE(E_A))=\tr(\cE^\dagger(F_B)E_A)
\end{align*}
for all $E_A\in\cL(\cH_A)$ and $F_B\in\cL(\cH_B)$.
Note that $(\cE^\dagger)^\dagger=\cE$, $\id_A^\dagger=\id_A$ and $(\cE\otimes\cF)^\dagger=\cE^\dagger\otimes\cF^\dagger$ for two maps $\cE$ and $\cF$.  Two easily verified properties which follow directly from this definition are 
\begin{align}\label{eq:unitalitytracepreservation}
\cE\textrm{ is unital if and only if }\cE^\dagger\textrm{ is
  trace-preserving}\ ,\ 
\end{align}
and
\begin{align*}
\cE\textrm{ is positive if and only if }\cE^\dagger\textrm{ is positive}\ .
\end{align*}
In particular, the last statement implies that 
\begin{align}\label{eq:completepositivitypreservation}
\cE\textrm{ is completely positive (CP) if and only if }\cE^\dagger\textrm{ is CP}\ .
\end{align}

Statements~\eqref{eq:unitalitytracepreservation} and~\eqref{eq:completepositivitypreservation} can be summarized as follows. Let us define $\CPTPM(\cH_A,\cH_B)$ as the set of quantum operations $\cE:\cL(\cH_A)\rightarrow\cL(\cH_B)$ and $\CPUM(\cH_B,\cH_A)$ as the set of completely positive unital maps $\cF:\cL(\cH_B)\rightarrow\cL(\cH_A)$.
We then have
\begin{lemma}\label{lem:adjointmapcharacterisation}
The adjoint map
\[
\dagger:\CPTPM(\cH_A,\cH_B)\rightarrow \CPUM(\cH_B,\cH_A)\]
is a bijection with inverse 
\[
\dagger:\CPUM(\cH_B,\cH_A)\rightarrow \CPTPM(\cH_A,\cH_B)\ .
\]
\end{lemma}
Let $d_A$ be the dimension of $\cH_A$ and 
let $\{\ket{x}_A\}_{x\in [d_A]}$ be an orthonormal basis of $\cH_A$. (We will restrict our attention to finite-dimensional Hilbert spaces.) Let $\cH_{A'}\cong\cH_{A}$ be a Hilbert space with orthonormal basis $\{\ket{x}_{A'}\}_{x\in[d_A]}$. The {\em maximally entangled state} on $\cH_A\otimes\cH_{A'}$ is defined as
\begin{align}
\ket{\Phi_{AA'}}=\frac{1}{\sqrt{d_A}} \sum_{x\in [d_A]} \ket{x}_{A} \otimes \ket{x}_{A'}\ .\label{eq:maximallyentangledstatedef}
\end{align}
The {\em Choi-Jamio\l{}kowski-map} $J$ takes operations
\mbox{$\cE:\cL(\cH_{A'})\rightarrow\cL(\cH_B)$} to operators
$J(\cE)\in\cL(\cH_{A}\otimes\cH_B)$. It is defined as
\begin{align*}
J(\cE)=d_A(\id_{A}\otimes\cE)(\proj{\Phi_{AA'}})\ .
\end{align*}
It has the following well-known properties. The equivalence of
statements~\eqref{it:CPTPMjamil} and~\eqref{it:CPUMjamil} in the
following lemma is an immediate consequence of
Lemma~\ref{lem:adjointmapcharacterisation}.

\begin{lemma}[Choi-Jamio\l{}kowski isomorphism,~\cite{jamiolkowski}]
\label{lem:jamiolkowski}
Let $\cH_A\cong \cH_{A'}$ and $\cH_B$ be arbitrary Hilbert spaces. 
The map $J$ bijectively maps
\begin{enumerate}[(i)]
\item\label{it:CPTPMjamil} the set $\CPTPM(\cH_{A'},\cH_B)$ to the set
  of operators $F_{AB}\geq 0$ with $\tr_B F_{AB}=\id_{A}$.
\item\label{it:CPUMjamil} the set $\CPUM(\cH_{A'},\cH_B)$ to the set
  of operators \mbox{$E_{AB}\geq 0$} with $\tr_A E_{AB}=\id_{B}$.
\end{enumerate}
\end{lemma}

Another concept we will need is the notion of {\em classicality},
which allows us to treat ensembles as quantum states. We will say that
a Hermitian operator $E_{AB}$ on a bipartite Hilbert space
$\cH_A\otimes\cH_B$ is {\em classical relative to an orthonormal basis
  $\{\ket{x}_A\}_{x\in [d_A]}$ of $\cH_A$} if it is a linear
combination of operators of the form $\proj{x} \otimes E_B$, where $x
\in [d_A]$ and $E_B$ is a Hermitian operator on~$\cH_B$.

\section{Main results and their derivation\label{sec:mainresultproof}}

We are now ready to prove our main statements.  We first focus on the
min-entropy in Section~\ref{sec:hminproof}. The interpretation of
max-entropy will be derived in Section~\ref{sec:proofhmax}.

\subsection{Proof of the operational characterization of
  $\hmin$\label{sec:hminproof}}

With Lemma~\ref{lem:slater} from Section~\ref{sec:semidefinite}, it is
straightforward to prove the following statement. Note that we
restrict our attention to finite-dimensional Hilbert spaces. Since the
optimizations are now taken over compact sets, we can replace $\inf$
and $\sup$ by $\min$ and $\max$, respectively.

\begin{lemma}\label{lem:operatorduality}
  Let $\cH_A$ and $\cH_B$ be finite-dimensional Hilbert spaces, and
  let $\rho_{AB}$ and $\sigma_B$ be nonnegative operators on
  $\cH_A\otimes\cH_B$ and $\cH_B$, respectively. Then
\begin{align}
\min_{\substack{\sigma_B\geq 0\\
\id_A\otimes\sigma_B\geq \rho_{AB}}} \tr(\sigma_B)=\max_{\substack{E_{AB}\geq 0\\
\tr_A(E_{AB})=\id_B
}}\tr\left(\rho_{AB}E_{AB}\right)\ .\label{eq:dualityapplied}
\end{align}
In addition, if $\rho_{AB}$ is classical on $\cH_A$
relative to an orthonormal basis $\{\ket{x}\}_x$, then the
maximization rhs.\ of~\eqref{eq:dualityapplied} can be further
restricted to those operators $E_{AB}$ which are classical on~$\cH_A$
relative to $\{\ket{x}\}_x$.
\end{lemma}

\begin{proof}
For a nonnegative operator~$E_{AB}$ with \mbox{$\tr_{A}E_{AB}\leq \id_B$}, we can
define the operator
\[
E'_{AB}=E_{AB}+\kappa_A \otimes
(\id_B-\tr_A(E_{AB}))\ ,
\] where $\kappa_A$ is an arbitrary normalized density
operator on $\cH_A$. We then have
\begin{align*}
\tr(\rho_{AB}E_{AB}')\geq \tr(\rho_{AB}E_{AB})
\end{align*}
 with $E_{AB}'\geq 0$ and  $\tr_A(E'_{AB})=\id_B$.
This shows that we can extend the maximization on the rhs.\ of~\eqref{eq:dualityapplied} to all operators $E_{AB}$ whose partial
trace $\tr_A(E_{AB})$ is bounded by $\id_B$ (instead of being equal to
$\id_B$). The claim is therefore equivalent to
\begin{align}\label{eq:dualityclaimrephrased}
\min_{\substack{\sigma_B\geq 0\\
\id_A\otimes\sigma_B\geq \rho_{AB}}} \tr(\sigma_B)=\max_{\substack{E_{AB}\geq 0\\
\tr_A(E_{AB})\leq \id_B
}}\tr(\rho_{AB}E_{AB})
\end{align}
To relate this to the general linear programming problem~\eqref{eq:primalanddual}, we define $\cV_1=\Herm(\cH_B)$ and $\cV_2=\Herm(\cH_A\otimes\cH_B)$ as the (real) vector spaces of Hermitian operators on $\cH_B$ and $\cH_A\otimes\cH_B$, respectively, with standard Hilbert-Schmidt inner product.
Furthermore,
we define the convex cones $K_1$ and $K_2$ as the set of nonnegative
operators in $\Herm(\cH_B)$ and $\Herm(\cH_A \otimes \cH_B)$, respectively. We claim that these cones are self-dual, i.e., $K_1^*=K_1$ and $K_2^*=K_2$.  This is easily seen from the spectral decomposition of a Hermitian
operator. Finally, we define $E:\cV_1\rightarrow\cV_2$ as the linear map
$E(\theta_B)=E\theta_B \assign \id_A\otimes\theta_B$. It is easy
to check that the adjoint $E^*:\cV_2\rightarrow\cV_1$ is equal to the
partial trace $\tr_A:\Herm(\cH_A\otimes\cH_B)\rightarrow\Herm(\cH_B)$;
indeed, for all $\theta_B\in\Herm(\cH_B)$ and
$F_{AB}\in\Herm(\cH_A\otimes\cH_B)$, we have
\begin{align*}
\spr{E\theta_B}{F_{AB}}_2&=\spr{\id_A\otimes \theta_B}{F_{AB}}_2\\
&=\tr((\id_A\otimes\theta_B)F_{AB})\\
&=\tr(\theta_B\tr_A(F_{AB}))\\
&=\spr{\theta_B}{\tr_A(F_{AB})}_1\ .
\end{align*}
We also set $b=\rho_{AB}$ and $c=\id_B$. With these definitions, we conclude that the two optimization problems defined by~\eqref{eq:dualityclaimrephrased} are a special instance of~\eqref{eq:primalanddual}; the claim is equivalent to the statement that the duality gap vanishes. According to Lemma~\ref{lem:slater}, it suffices to check Slater's interiority condition. For this purpose, we set $v=2\lambda_{\max}(\rho_{AB})\cdot\id_B$, where $\lambda_{\max}$ denotes the maximal eigenvalue. Clearly, $v$ is in the interior of~$K_1$. We also have
\[
E v-b=2\lambda_{\max}(\rho_{AB})\id_{AB}-\rho_{AB}>0\ ,
\]  
hence $E v-b\in\interior K_2$; this proves the claim~\eqref{eq:dualityclaimrephrased}.

To prove the claim about the case where  $\rho_{AB}$ is classical relative to an orthonormal basis $\{\ket{x}\}_{x}$ of
$\cH_A$, we simply set $\cV_2=\myspan\{\proj{x}\}_x\otimes
\Herm(\cH_B)$ equal to the set of Hermitian operators that are
classical on $\cH_A$. The remainder of the proof
is identical to the general case.
\end{proof}
Observe that the lhs.\ of~\eqref{eq:dualityapplied} is equivalent to a
minimization of the distance measure $\RelEntr$
from~\eqref{eq:Dinfty}, i.e., we have
\begin{align}
\log \min_{\substack{\sigma_B\geq 0\nonumber\\
\id_A\otimes\sigma_B\geq \rho_{AB}}} \tr(\sigma_B)&=\min_{\substack{\sigma_B\geq 0\\ \tr(\sigma_B)=1}} \RelEntr(\rho_{AB}||\id_A\otimes\sigma_B)\\
&=-\hmin(A|B)_\rho\ .\label{eq:hminrewritten}
\end{align}
Let us discuss the case where $\rho_{XB}$ is classical on $X$. Lemma~\ref{lem:operatorduality} allows us to show that the min-entropy $\hmin(X|B)_\rho$ is equivalent to the ``guessing-entropy'' of $X$ given $B$.
\begin{theorem}\label{lem:hguess}
Let $\rho_{X B}=\sum_{x}p_x\proj{x}\otimes\rho^x_B$ be classical on~$\cH_X$. Then 
\begin{align*}
\hmin(X|B)_\rho=-\log \, \pguess(X|B)_\rho\ ,
\end{align*}
where  $\pguess(X|B)_\rho$ is the maximal probability of decoding $X$ from $B$ with a POVM $\{E^x_B\}_{x}$ on $\cH_B$, i.e.,
\begin{align*}
\pguess(X|B)_\rho &\assign \!\! \max_{\{E^x_B\}_x} \sum_{x} p_x\tr(E^x_B\rho^x_B)\ .
\end{align*}
\end{theorem}
\begin{proof}
According to~\eqref{eq:hminrewritten}, it suffices to show that  the rhs.\ of~\eqref{eq:dualityapplied} is equal to $\pguess(X|B)_\rho$. But this is a direct consequence of the fact that every nonnegative operator $E_{X B}$ with $\tr_X(E_{X B})=\id_B$ which is classical on $\cH_X$ has the form
\begin{align*}
E_{X B}=\sum_{x}\proj{x}\otimes E^x_B\ ,
\end{align*}
where the family $\{E^x_B\}_x$ is a POVM on $\cH_B$.
\end{proof}

The Choi-Jamio\l{}kowski isomorphism yields an operational
interpretation of the min-entropy in the general case.  We can express the
min-entropy as the maximal achievable singlet fraction as follows.

\begin{theorem}\label{thm:singletfraction}
The min-entropy of a state $\rho_{AB}$ on $\cH_A\otimes\cH_B$ can be expressed as
\begin{align}\label{eq:mainclaimhmin}
\hmin(A|B)_\rho&=-\log \pcorr(A|B)_{\rho} \ ,
\end{align}
where $\pcorr(A|B)_\rho$ is the maximal achievable singlet fraction, i.e., 
\begin{align*} 
 \pcorr(A|B)_{\rho} \assign d_A \max_{\cF} F\bigl((\id_A \otimes \cF)(\rho_{A
   B}), \proj{\Phi_{AA'}}\bigr)^2\ ,
\end{align*}
with maximum taken over all  quantum operations~\mbox{$\cF:\cL(\cH_B)\rightarrow \cL(\cH_{A'})$}, \mbox{$\cH_{A'}\cong \cH_A$} and $\ket{\Phi_{AA'}}$ defined by~\eqref{eq:maximallyentangledstatedef}.
\end{theorem}

\begin{proof}
Let us rewrite statement~\eqref{eq:mainclaimhmin} as
\begin{align}
\begin{split} \label{eq:mainclaimdminmodified} \raisetag{8ex}
&\min_{\substack{\sigma_B\geq 0\\ \tr(\sigma_B)=1}}
  \RelEntr(\rho_{AB}||\id_A\otimes\sigma_B) \\
&\quad = \log \left(d_A\cdot
\max_{\cF} F\bigl((\id_A \otimes \cF)(\rho_{A
   B}), \proj{\Phi_{AA'}}\bigr)^2     \right)\  ,
\end{split}
\end{align}
where $\ket{\Phi_{AA'}}$ is the maximally entangled state. Let
$E_{AB}$ be a nonnegative operator on $\cH_A\otimes\cH_B$ with $\tr_A
E_{AB}=\id_B$, and let \mbox{$\cE=J^{-1}(E_{AB})\in
  \CPUM(\cH_{A'},\cH_B)$} be the unital map corresponding to $E_{AB}$
under the Choi-Jamio\l{}kowsi isomorphism
(cf.~Lemma~\ref{lem:jamiolkowski}~\eqref{it:CPUMjamil}).  Let
\mbox{$\cF=\cE^\dagger:\cL(\cH_B)\rightarrow\cL(\cH_{A'})$} be the
adjoint quantum operation
(cf.~Lemma~\ref{lem:adjointmapcharacterisation}).  By definition of
$\cE$ and the adjoint
\mbox{$(\id_A\otimes\cE)^\dagger=\id_A\otimes\cE^\dagger=\id_A\otimes\cF$}
we have
\begin{align*}
\tr(\rho_{AB}E_{AB})&=d_A\tr\left(\rho_{AB}(\id_A\otimes\cE)(\proj{\Phi_{AA'}})\right)\\
&=d_A\tr\left((\id_A\otimes\cE)^\dagger (\rho_{AB})\proj{\Phi_{AA'}}\right)\\
&=d_A\tr\left((\id_A\otimes\cF)(\rho_{AB})\proj{\Phi_{AA'}}\right)\ .
\end{align*} 
Observe that the operators $E_{AB}\geq 0$ with $\tr_A E_{AB}=\id_B$
are in one-to-one correspondence with quantum operations $\cF\in
\CPTPM(\cH_B,\cH_{A'})$ constructed in this fashion. The
claim~\eqref{eq:mainclaimdminmodified} therefore follows from
Lemma~\ref{lem:operatorduality} and~\eqref{eq:hminrewritten}.
\end{proof}

\begin{remark} \label{rem:singlfractionmax} The result of
  Theorem~\ref{thm:singletfraction} can be extended to give an
  alternative expression for the maximal achievable fidelity
   with a non-maximally entangled state
  \mbox{$\ket{\Psi_{AA'}}=\sum_\lambda
  \sqrt{\lambda}\ket{\lambda}_A\ket{\lambda}_{A'}\in\cH_A\otimes\cH_{A'}$}. We
  assume that \mbox{$\cH_{A'}\cong\cH_A$} and that $\ket{\Psi_{AA'}}$
  has maximal Schmidt rank. Let $\tau_A=\tr_{A'}\proj{\Psi_{AA'}}$ be
  its reduced density operator.  Then
\begin{align}
\begin{split}  \label{eq:mainclaimdmin} \raisetag{8ex}
&\min_{\substack{\sigma_B\geq 0\\ \tr(\sigma_B)=1}}
  \RelEntr(\rho_{AB}||\tau_A^{-1}\otimes\sigma_B)\\
&\qquad=\log \max_{\cF} F\bigl((\id_A \otimes \cF)(\rho_{A
   B}), \proj{\Psi_{AA'}}\bigr)^2
\end{split}
\end{align}
for any bipartite state $\rho_{AB}$ on
$\cH_A\otimes\cH_{B}$. Statement~\eqref{eq:mainclaimdmin} follows by
substituting
\mbox{$(\tau_A^\half\otimes\id_B)\rho_{AB}(\tau_A^\half\otimes\id_B)$}
for~$\rho_{AB}$ in~\eqref{eq:mainclaimdminmodified}, using the fact
that conjugating with an invertible matrix does not change operator
inequalities, and~\mbox{$\sqrt{d_A}\cdot
  (\tau_A^\half\otimes\id_B)\ket{\Phi_{AA'}}=\ket{\Psi_{AA'}}$}.
\end{remark}

\subsection{Proof of the operational characterization of $\hmax$\label{sec:proofhmax}} 
To obtain the operational characterization of $\hmax$, we use
Theorem~\ref{thm:singletfraction}. Recall the definition of the
decoupling accuracy of a bipartite state $\rho=\rho_{AB}$, that is,
\begin{align*}
 \pdecpl(A|B)_{\rho} \assign d_A \max_{\sigma_B} F(\rho_{AB},\tau_A\otimes\sigma_B)^2\ ,
\end{align*}
where  $d_A$ is the dimension of~$\cH_A$ and $\tau_A$ is the
completely mixed state on $\cH_A$. We begin by showing the following
lower bound on the decoupling accuracy.
\begin{lemma}
For all bipartite states $\rho_{AB}$, we have 
\begin{align*}
2^{\hmax(A|B)_\rho}&\leq   \pdecpl(A|B)_{\rho}\ .
\end{align*}
\end{lemma}
\begin{proof}
Let $\rho_{ABC}=\proj{\varphi_{ABC}}$ be a purification of $\rho_{AB}$, and let $\cF:\cL(\cH_C)\rightarrow \cL(\cH_{A'})$ be a quantum operation that satisfies (cf.~Theorem~\ref{thm:singletfraction})
\begin{align*}
2^{-\hmin(A|C)}=d_A F((\id_A\otimes\cF)(\rho_{AC}),\ket{\Phi_{AA'}})^2\ .
\end{align*} 
Let \mbox{$\rho'_{A A' B R}=\proj{\varphi'_{A A' B R}}$} be a
purification of \mbox{$\rho'_{A B A'}=(\id_{AB}\otimes\cF)(\rho_{ABC})$}. We then have 
\begin{align}
2^{\hmax(A|B)}&=2^{-\hmin(A|C)}\nonumber\\
&=d_A F(\rho'_{AA'}, \ket{\Phi_{AA'}})^2\label{eq:fidelityfmaxs}\
\end{align}
However,
\begin{align*}
F( \rho'_{AA'}, \ket{\Phi_{AA'}})&=F(\ket{\varphi'_{A A' B R}}, \ket{\Phi_{AA'}}\otimes\ket{\theta_{BR}})
\end{align*}
for some state $\ket{\theta_{BR}}$ on $\cH_B\otimes\cH_R$. By the monotonicity of the fidelity, we therefore get
\begin{align*}
F(\rho'_{AA'},\ket{\Phi_{AA'}})&\leq F(\rho_{AB},\tau_A\otimes\tr_R\proj{\theta_{BR}})\\
&\leq \max_{\sigma_B}  F(\rho_{AB},\tau_A\otimes\sigma_B)\ ,
\end{align*}
where we used the fact that $\rho'_{AB}=\rho_{AB}$. Inserting this into~\eqref{eq:fidelityfmaxs} gives the claim.
\end{proof}
The proof of the converse inequality closely follows a derivation in~\cite{horodeckietal08}. We include it here for completeness.

\begin{lemma}
For all bipartite states $\rho_{AB}$, we have 
\begin{align*}
2^{\hmax(A|B)_\rho}&\geq   \pdecpl(A|B)_{\rho}\ .
\end{align*}
\end{lemma}
\begin{proof}

We use the following fact, which is a consequence of the fact that all purifications of a fixed state are related by a unitary transformation on a (possibly extended) ancilla. If $\ket{\phi_{A B C C'}}$ has a reduced state of the form $\tr_{CC'} \proj{\phi_{A B C C'}}=\tau_A\otimes\sigma_B$, where $\tau_A$ is the completely mixed state on $\cH_A$, then there exists a unitary $U_{CC'}$ such that 
\begin{align}
(\id_{AB}\otimes U_{CC'})\ket{\phi_{A B C C'}}=\ket{\Phi_{AC}}\ket{\theta_{BC'}}\label{eq:decouplingprod}
\end{align} for some state $\ket{\theta_{BC'}}$ on $\cH_B\otimes\cH_{C'}$, where $\ket{\Phi_{AC}}$ denotes the fully entangled state on $\cH_A\otimes\cH_C$ (without loss of generality, we can assume that $d_A\leq d_C$).

Let $\sigma_B$ be an arbitrary density matrix on $\cH_B$. Let \mbox{$\rho_{ABC}=\proj{\psi_{ABC}}$} be a purification of $\rho_{AB}$, where we assume the dimension of $\cH_C$ to be sufficiently large. 

According to the definition of the fidelity, there exists a purification $\ket{\phi_{A B C C'}}$ of $\tau_A\otimes\sigma_B$ such that
\begin{align*}
F(\rho_{AB},\tau_A\otimes\sigma_B) &=F(\ket{\psi_{ABC}}\ket{0_{C'}},\ket{\phi_{A B C C'}})\ .
\end{align*}
Applying the unitary $U_{CC'}$ from~\eqref{eq:decouplingprod} gives 
\begin{align*}
F(\rho_{AB},\tau_A\otimes\sigma_B)&=F(\ket{\psi'_{A B C C'}},\ket{\Phi_{AC}}\ket{\theta_{BC'}})\ ,
\end{align*} 
where $\ket{\psi'_{A B C C'}}=(\id_{AB}\otimes U_{CC'}) \ket{\psi_{ABC}}{\ket{0_{C'}}}$
because of the invariance of the fidelity under unitary operations. Using the monotonicity of the fidelity, we conclude that
\begin{align*}
F(\rho_{AB},\tau_A\otimes\sigma_B)&\leq F(\tr_{BC'}\proj{\psi'_{A B C C'}},\ket{\Phi_{AC}})\\
&=F((\id_A\otimes\cF)(\rho_{AC}),\ket{\Phi_{AC}})\ ,
\end{align*}
where $\cF:\cL(\cH_C)\rightarrow\cL(\cH_C)$ is the quantum operation
$\cF(\rho)=\tr_{C'}(U_{CC'}(\rho\otimes\proj{0}_{C'})U_{CC'}^\dagger)$. Squaring
both sides of the previous inequality, multiplying by $d_A$, taking
the maximum over all quantum operations and using 
Theorem~\ref{thm:singletfraction} therefore gives
\begin{align*}
d_A F(\rho_{AB},\tau_A\otimes\sigma_B)^2&\leq  2^{-\hmin(A|C)_{\rho}}\ .
\end{align*}
Since $\sigma_B$ was arbitrary, we can maximize the lhs.\ over all
$\sigma_B$. The claim then follows from the definitions of
$\pdecpl(A|B)_\rho$ and~$\hmax(A|B)_\rho$.
\end{proof}

In summary, we have shown the following result.
\begin{theorem}\label{thm:hmaxinterpretation}
Let $\rho_{AB}$ be a state on $\cH_A\otimes\cH_B$, and let $\tau_A$ be the completely mixed state on $\cH_A$. Then
\begin{align*}
\hmax(A|B)_\rho&=\log \pdecpl(A|B)_\rho\ ,
\end{align*}
where $\pdecpl(A|B)_{\rho}$ is the decoupling accuracy, defined by
\begin{align*}
\pdecpl(A|B)_{\rho} \assign d_A \max_{\sigma_B} F(\rho_{AB},\tau_A\otimes\sigma_B)^2\ ,
\end{align*}
with the maximum taken over all normalized states $\sigma_B$ on $\cH_B$.
\end{theorem} 

\section{Conclusions} \label{sec:conclusions} 

\begingroup
\begin{samepage}
\begin{table*}[t!]
\begin{center}
\begin{tabular}{c|c|c|}
  goal & quality of a state $\rho_{A B}$ & amount of extremal states  \\
  (extremal state) &  (measured in terms of overlap) & contained in a state $\rho_{A B}$ \\
  & & (measured in \# of qubits on $A$) \\[1.4ex] \hline & & \\[-0.4ex]
  \begin{minipage}{5.5cm}\vspace{1mm} $A$ fully entangled with $B$ \\[1.6ex]
    classical $A$ fully determined by $B$ \end{minipage}  
  & $ \left. \begin{matrix}  -\log \pcorr(A|B)  \\[1ex] - \log
      \pguess(A|B) \end{matrix} \right\} = \hmin(A|B)$ &   $-
  \ellmerg^\eps(A|B) \stackrel{\eqref{eq:ellmerg}}{\approx}
  -\hmax^{\eps'}(A|B)$
\\[-0.5ex] & & \\[1ex] \hline & & \\[-0.5ex]
  \begin{minipage}{5.5cm}\vspace{1mm} $A$ fully mixed and indep.\ of $B$ \\[1.5ex]
    classical $A$ uniform and indep.\ of $B$ \end{minipage} & 
$\left. \begin{matrix}  \log \pdecpl(A|B) \\[1ex] \log
    \psecr(A|B) \end{matrix} \right\} = \hmax(A|B)$ & $\elldec^\eps(A|B) \stackrel{\eqref{eq:elldecpl}}{\approx}
\hmin^{\eps'}(A|B)$ \\[-0.5ex] & & \\[1ex]
\end{tabular}
\end{center}
\caption{Operational interpretations of (smooth) min- and
  max-entropies. The approximation ($\approx$) indicates that equality
  holds up to an additive term of order $\log \frac{1}{\eps}$ and for
  an appropriate choice of the smoothness parameter
  $\eps'$. \label{tb:inter}}
\end{table*}
\end{samepage}
\endgroup

In information theory, entropies are generally interpreted as
\emph{measures of uncertainty}. One method to make this interpretation
more precise is to establish relations between entropy measures and
\emph{operational quantities}, that is, quantities that characterize
actual information-theoretic tasks. 

Here we consider a general scenario consisting of a (possibly
quantum-mechanical) system $A$ as well as an observer with (quantum or
classical) \emph{side information} $B$. The uncertainty of the
observer about the state of system $A$ then depends on the
distribution of these states as well as the correlation between $A$
and $B$.

There are two extreme situations, namely when $A$ is completely
undetermined and when $A$ is determined. Taking into account the side
information $B$, these two situations are described as follows.
\begin{enumerate}
\item[1.] \label{it:fullcor} The state of $A$ is fully correlated with
  (parts of) $B$.\footnote{In the general case where $A$ and $B$ are
    quantum-mechanical systems, full correlation is akin to maximal  entanglement.}
\item[2.] The state of $A$ is uniformly distributed and independent of the
  side information $B$.
\end{enumerate}
Note that in the first case, the requirement is merely that~$A$ is correlated with \emph{parts} of $B$. This is because the
side-information $B$ may consist of additional information that is
unrelated to $A$.

For any given state $\rho_{A B}$, we may characterize the uncertainty
of $A$ given $B$ by the distance to these extreme situations. If we
take as a distance measure the \emph{overlap} (i.e., the square of the
\emph{fidelity}), we retrieve the definitions of $\pcorr(A|B)$ and
$\pdecpl(A|B)$ (see~\eqref{eq:pcorrdef} and~\eqref{eq:pdecpldef},
respectively). Our main results imply that these correspond to
$\hmin(A|B)$ and $\hmax(A|B)$, respectively. We thus conclude that
$\hmin(A|B)$ quantifies the closeness to a situation where $A$ is
determined by $B$, and, likewise, $\hmax(A|B)$ corresponds to the
closeness to a situation where $A$ is independent of $B$ (see second
column of Table~\ref{tb:inter}).

Given a bipartite state $\rho_{A B}$, we may also ask for the number of maximally entangled or completely independent qubits one can extract from~$A$. Very roughly speaking, this is the
idea underlying the definitions of $\ellmerg^\eps(A|B)$ and $\elldec^\eps(A|B)$,
respectively (see~Section~\ref{sec:opint} for more details, in
particular the interpretation of negative quantities). Remarkably,
these quantities are (approximately) given by the smooth entropies
$\hmax^{\eps'}(A|B)$ and $\hmin^{\eps'}(A|B)$ (see last column of
Table~\ref{tb:inter}).\footnote{Note that compared to the discussion of the distance, the role of $\max$ and $\min$ is interchanged.}

Despite these similarities between the (previously known) operational
interpretations summarized in the last column of
Table~\ref{tb:inter} and those given in the second column (the ones
derived here), there are at least two fundamental differences. The
first is that the new interpretations are exact and, in particular,
valid without a smoothness parameter. In contrast, all previously
established interpretations only hold up to additive terms of the
order $\log \frac{1}{\eps}$, where $\eps$ is a smoothness parameter
(whose meaning is that of an error or failure probability). A second
difference is that there does not seem to exist an obvious asymptotic
counterpart for our identities. In particular, there are no analogous
operational interpretations of the von Neumann entropy.

The results of this paper suggest that studying operationally defined
quantities may be a viable approach to identifying relevant
single-shot information measures in a multipartite setting. Of
particular interest is the conditional mutual information, which has
only recently been given an asymptotic
interpretation~\cite{DevYard06}.

\subsubsection*{Acknowledgments}
RK acknowledges support by NSF grants PHY-0456720 and PHY-0803371. CS is supported by EU fifth
framework project QAP IST 015848 and the NWO VICI project 2004-2009. 
RR acknowledges support from the Swiss National Science Foundation (grant no.~200021-119868).
\bigskip
\vfill{}

\pagebreak
\bibliographystyle{IEEEtran}


\end{document}